\documentclass[fleqn,usenatbib]{mnras}

\usepackage{newtxtext,newtxmath} 
\usepackage{CJKutf8}
\usepackage[T1]{fontenc}
\usepackage{ae,aecompl}
\usepackage[caption=false]{subfig}
\usepackage{graphicx}	
\usepackage{amsmath}	
\usepackage{graphicx}	
\usepackage{float,lscape}
\usepackage{comment}
\usepackage[utf8]{inputenc}
\usepackage[dvipsnames]{xcolor}
\usepackage{siunitx}
\usepackage{verbatim}
\usepackage[english]{babel}
\usepackage[T1]{fontenc}
\usepackage{ae,aecompl}
\usepackage{textcomp}
\usepackage[normalem]{ulem}
\usepackage{soul}
\usepackage{lipsum}
\usepackage{booktabs}
\usepackage{url}
\usepackage{threeparttable}

\usepackage[inline]{enumitem}

\newcommand*{\email}[1]{\href{mailto:#1}{\nolinkurl{#1}} } 

\numberwithin{equation}{section}

\title{X-ray constraint for the unseen companion of V723 Mon: it is a mass-gap black hole rather than binary neutron stars}

\author[Y.~Li et al.]{\parbox{\textwidth}{
Yan~Li$^{1,2,}\footnotemark[1]$
  \begin{CJK*}{UTF8}{gbsn}
    (李彦),
  \end{CJK*} 
Erlin~Qiao$^{3,4}$ 
  \begin{CJK*}{UTF8}{gbsn}
    (乔二林),
  \end{CJK*} 
and Rong-Feng~Shen$^{1,2}\footnotemark[2]$ 
  \begin{CJK*}{UTF8}{gbsn}
    (申荣锋)
  \end{CJK*} 
}
\vspace{0.4cm}\\  
\parbox{\textwidth}{
$^{1}$ School of Physics and Astronomy, Sun Yat-Sen University, Zhuhai, 519082, P. R. China\\
$^{2}$ CSST Science Center for the Guangdong-Hongkong-Macau Greater Bay Area, Sun Yat-Sen University, Zhuhai, 519082, P. R. China \\
$^{3}$ Key Laboratory of Space Astronomy and Technology, National Astronomical Observatories, Chinese Academy of Sciences, Beijing, 100101, P. R. China \\
$^{4}$ School of Astronomy and Space Sciences, University of Chinese Academy of Sciences, 19A Yuquan Road, Beijing, 100049, P. R. China\\
}}

\date{Accepted XXX. Received YYY; in original form ZZZ}

\pubyear{2022}

\usepackage{newtxtext,newtxmath}

\begin{document}
\newcommand{\de}{\mathrm d}
\label{firstpage}
\pagerange{\pageref{firstpage}--\pageref{lastpage}}
\maketitle

\begin{abstract}
Recently, the red giant V723 Mon is reported to have an unseen companion with a mass of $3.04\pm0.06M_{\odot}$, but question remains about whether it is a single (thus the so-called mass-gap) black hole or an inner binary of two more ordinary compact objects (neutron stars or white dwarfs). In this work, we estimate the X-ray emission by considering the wind-fed accretion from V723 Mon onto the compact companion. We analyze three different scenarios of the dark companion, i.e., a single black hole, binary neutron stars and binary of a neutron star and a white dwarf. We show that the single black hole is the most favored scenario. We also calculate the synchrotron emission from the bow shock caused by the interaction of the compact companion with the wind. We find this emission peaks at $\sim$ 0.1-1 GHz, with a flux density $\sim$ 1 mJy, which is expected to be detected by observations with higher angular resolution in the future. \\\\

\end{abstract}


\section{Introduction}\label{sec:introduction}

\renewcommand{\thefootnote}{\fnsymbol{footnote}}
\footnotetext[1]{E-mail: \email{liyan287@mail2.sysu.edu.cn}}
\footnotetext[2]{E-mail: \email{shenrf3@mail.sysu.edu.cn}}
\renewcommand{\thefootnote}{\arabic{footnote}}

Due to the lack of low-mass black holes (BHs) in the observed sample of Galactic X-ray binaries, it has been empirically suggested that a mass gap in the range of $\sim2.5-5M_{\odot}$ could exist, separating neutron stars (NSs) from BHs \citep{bailyn98,ozel10,farr11}. The existence of the mass gap is not a definitive fact. Recently, \cite{thompson19} reported the discovery of a BH of $3.3^{+2.8}_{-0.7}$ $M_{\odot}$ in a noninteracting binary system with a red giant. Furthermore, a compact object with a mass of $2.59^{+0.08}_{-0.09}$ $M_{\odot}$ is reported in the gravitational-wave observation of GW190814, which is either the heaviest neutron star or the lightest black hole ever observed \citep{abbott20}.   

The discoveries of those mass-gap objects lead to important astrophysical implications. If a BH or an NS forms through a supernova (SN) explosion as a result of single-star evolution, the progenitor star must be massive ($\geq8M_{\odot}$) \citep{maximiliano21}. \cite{belczynski12} show that, the rapid SN explosion model in \cite{fryer12}, which is driven by rapidly growing instabilities, disfavors the presence of the mass-gap BHs, while the delayed-explosion model in \cite{fryer12} favors the continuous distribution of BH mass. Therefore, the existence of the mass-gap object may inform us more about the fate of the massive star and shed light on the SN explosion mechanism. Besides, a mass-gap object may be produced as the remnant of a binary neutron star merger, whose nature is dependent on the upper mass limit of a stable NS \citep{abbott20}, which inturn is determined by the unknown NS equation of state (EOS). Thus identifying new mass-gap objects is pivotal in understanding the NS EOS, and the BH formation channel.

A potential mass-gap object of $3.04\pm0.06M_{\odot}$ is reported as the dark companion of V723 Mon, a nearby, bright red giant (RG) in the constellation of Monoceros \citep{jaya21}. The radius of the RG is less than that of its Roche lobe \citep{jaya21}, implying that the V723 Mon is detached from its dark companion. In this case, the RG's stellar wind is accreted by the compact object. In this wind-fed accretion, the accretion geometry around the compact accretor takes an almost spherical Bondi flow \citep{hoyle39,bondi44}, accompanied by a bow shock around the accretor \citep{hunt71,illarionov75}. 

The X-ray emission from the accretion onto a compact object, i.e., BH, NS and white dwarf (WD), is a useful probe to the nature of the compact object. If the accreting gas cools efficiently, i.e., via radiative loss of the energy converted from the gravitational potential energy, the orbiting gas would form a standard thin disk-like configuration \citep{shakura73,lynden74}. For the thin disk model, the  radiative efficiency is simply determined by the compactness of the central compact object, i.e., the ratio of its mass over its radius \citep{frank02}. The larger the compactness is, the greater the efficiency is. Therefore, due to the difference in their compactness owned by BH, NS and WD, their emission properties (e.g., luminosity, hardness, etc.) differ. However, if the cooling of accreting gas is inefficient, the accretion material would be an advection dominated accretion flow (ADAF) \citep{narayan94,narayan95} and only a tiny fraction of energy is released through radiation. For the ADAF disk, the advected energy of the BH accretion disk is lost into the horizon, whereas the thermalized energy of the NS or WD accretion disk is eventually radiated from their stellar surface \citep{narayan95}. Therefore, the observation of the resulting X-ray luminosity is critical to distinguish different accretors.

The nature of the dark companion of V723 Mon is a subject of debate. If the companion is a single compact object, it would be a BH falling in the low mass gap \citep{jaya21}. \cite{masuda21} analyzed the tidal effects on the radial velocities (RV) of V723 Mon and concluded that there is no need for a third body to explain the periodic RV residuals. However, there remains other two possible scenarios that the compact companion is composed of binary neutron stars (BNS) or binary of an NS and a WD, both of which have not been ruled out yet \citep{jaya21}. 

In this paper, with an aim to reveal the nature of the dark compact companion of V723 Mon, we study the radiative signatures of the accretor in three different scenarios, i.e., a single BH, BNS and binary of an NS and a WD as the dark companion. In Section 2 we first derive the wind capture rate onto the compact accretor, then compare the estimated X-ray emission from the three scenarios with the observations of the V723 Mon system. Section 3 presents our estimate of the synchrotron radio emission from the bow shock, a signature of the wind-fed accretion in V723 Mon. In Section 4 we conclude the results and discuss the implications.

\section{wind-fed accretion onto the compact companion}\label{sec:sample}

The parameters of the V723 Mon from the analysis performed by \cite{jaya21} are summarized in Table \ref{tab:v723}. The stellar radius and the Roche limits of the RG are approximately equal to $25R_{\odot}$ and $30R_{\odot}$, respectively, suggesting that the RG is inside its Roche lobe. Therefore, the RG and its dark compact companion form a detached system. As the sizes of the two components are smaller than those of their critical Roche lobes, the accretion onto the compact object must come from the stellar wind from the RG \citep{illarionov75}.


\begin{table}
\centering
\caption{Parameters of the V723 Mon system \citep{jaya21}. Rows from top to bottom show companion mass, RG mass, RG radius, orbital inclination, orbital separation, distance, orbital period, RG luminosity, observed X-ray luminosity and orbital eccentricity.}
\label{tab:v723}%
\begin{threeparttable}
\begin{tabular}{p{3.9cm}p{2.7cm}}
	    \toprule
	    parameter & value \\
	    \midrule
	        \midrule
                \ $M_{\rm co}$  & \ $3.04\pm0.06$ $M_{\odot}$ \\
                \ $M_{\rm RG}$  & \ $1\pm0.07$ $M_{\odot}$ \\
                \ $R_{\rm RG}$  & \ $24.9\pm0.7$ $R_{\odot}$ \\
                \ $i$  &\ $87_{-1.4}^{+1.7}$ $^{\circ}$ \\
                \ $a\times \rm{sin}$$i$      & \ 102 $R_{\odot}$ \\
                \ $d$      &\ $460$ $\rm pc$ \\
                \ $P_{\rm orb}$    &\ $59.9$ $\rm day$ \\
                \ $L_{\rm RG}$  &\ $173$ $L_{\odot}$ \\
                \ $L_{\rm X,obs}$ $^{*}$  &\ $7.6\times {10^{29}}$ $\rm erg$ $\rm s^{-1}$\\
                \ $e_{\rm orb}$ &\ 0.015\\
              \bottomrule
\end{tabular}%
	\begin{tablenotes}
	\item[*] 0.3-2.0 keV.
	\end{tablenotes}
\end{threeparttable}	
 
\end{table}%

The wind mass-loss rate from the RG can be estimated as \citep{reimers75} 
\begin{equation}
\begin{split}
\dot M_{\rm w} &= 4 \times {10^{ - 13}}{\eta _{\rm R}} \left(\frac{L_{\rm RG}}{L_\odot} \right)\left(\frac{{R_{\rm RG}}}{R_{\odot}}\right)\left(\frac{M_{\rm RG}}{M_{\odot}}\right)^{-1}{M_ \odot }{\rm{y}}{{\rm{r}}^{ - 1}} \\
&\approx 8 \times {10^{ - 10}}{M_ \odot }{\rm{y}}{{\rm{r}}^{ - 1}},
\end{split}
\end{equation}
where $\eta_{\rm R}$ is set to 0.477 \citep{mcdonald15}, which accounts for the efficiency of the mass-loss rate from the RG. 

Part of the wind is gravitationally captured  by the compact object within a cylinder of radius ${R_{\rm cap}} = {{2GM_{\rm co}} \mathord{\left/
 {\vphantom {{2GM_{\rm co}} {v_r^2}}} \right.
 \kern-\nulldelimiterspace} {(v_{\rm rel}^2}+c_{\rm s}^2)}$, where $v_{\rm rel}=\sqrt{v_{\rm orb}^2+v_{\rm w}^2}$ is the relative velocity between the compact object and the stellar wind, $M_{\rm co}$ is the mass of the compact object,  $v_{\rm orb}$ is the orbital velocity of the compact object, $v_{\rm w}$ is the wind velocity, and $c_{\rm s}$ is the sound speed \citep{bondi52}. As $c_{\rm s}$ is negligible compared with $v_{\rm rel}$, we have
\begin{equation}
{R_{\rm cap}} = {{2GM_{\rm co}} \mathord{\left/
 {\vphantom {{2Gm} {v_r^2}}} \right.
 \kern-\nulldelimiterspace} {v_{\rm rel}^2}}.
\end{equation}

In the original Bondi-Hoyle-Lyttleton accretion scenario, the wind capture rate onto the compact companion can be estimated as ${\dot M_{\rm cap}} = \pi R_{\rm cap}^2{v_{\rm rel}}{\rho _{\rm w}}$ \citep{hoyle39}, where $\rho_{\rm w}=\dot{M}_{\rm w}/(4\pi a^2v_{\rm w})$ \citep{illarionov75} is the wind density. Then
\begin{equation}
{\dot M_{\rm cap}} = \frac{{\dot M}_{\rm w}}{4}\frac{{R_{\rm cap}^2}}{{a{}^2}}\frac{{{v_{\rm rel}}}}{{{v_{\rm w}}}},
\end{equation}
where $a$ is the separation between the RG and the compact object.

The wind velocity $v_{\rm w}$ is crucial to determine the capture rate. Assuming that the wind escapes radially from the RG, it is conveniently to express the wind velocity as a function of the distance $r$ from the RG  
\begin{equation}
v_{\rm w}=\alpha_{\rm w}(r)v_{\rm esc},
\end{equation}
where $v_{\rm esc}=\sqrt{2GM_{\rm RG}/R_{\rm RG}}$ is the escape velocity from the RG. For low-mass RG, the scaling factor $\alpha_{\rm w}(r)$ is given by \cite{yungelson19} as
\begin{equation} 
\begin{split}
{\alpha _{\rm w}}\left( r \right) = &\left( {1 - \frac{{5\,{{{\rm{km}}} \mathord{\left/
 {\vphantom {{{\rm{km}}} {\rm{s}}}} \right.
 \kern-\nulldelimiterspace} {\rm{s}}}}}{{{v_{\rm esc}}}}} \right)\exp \left\{ { - {{\left[ {\frac{1}{5}\left( {\frac{r}{{{R_{\rm RG}}}} - 1} \right)} \right]}^2}} \right\} \\
&+ \frac{{5\,{{{\rm{km}}} \mathord{\left/
 {\vphantom {{{\rm{km}}} {\rm{s}}}} \right.
 \kern-\nulldelimiterspace} {\rm{s}}}}}{{{v_{\rm esc}}}}.
\end{split}
\end{equation}
For the V723 Mon system, $r \sim a=102R_{\odot}$, thus, Eq.(2.5) gives $\alpha_{\rm w}$ $\approx$ 0.69.

Whether accreted material has sufficient angular momentum to form an accretion disk around the compact object is critical to determine the observed X-ray luminosity. Forming an accretion disk requires that the outer edge of the disk, $R_{\rm d}$, should be larger than the innermost physical radius of the accretor \citep{illarionov75,shapiro76}.

The change of the speed and density of the incident gas across the cylinder cross-section that intersects the accretor would result in a net angular momentum of the captured material, which \cite{shapiro76} derived to be $\ell_{\rm cap} \sim \frac{1}{2}R_{\rm cap}^2\Omega_{\rm orb}$, where $\Omega_{\rm orb}$ is the orbital angular velocity. $R_{\rm d}$ is where the specific Keplerian angular momentum equals $\ell_{\rm cap}$ \citep{shapiro76}, i.e., $\sqrt{GM_{\rm co}R_{\rm d}}=\frac{1}{2}R_{\rm cap}^2\Omega_{\rm orb}$, thus
\begin{equation}
R_{\rm d}=\frac{R_{\rm cap}^4\Omega_{\rm orb}^2}{4GM_{\rm co}}. 
\end{equation}

From the above and the V723 Mon parameters, we get $R_{\rm cap}\simeq 5.5\times 10^{12}$ cm, $\dot M_{\rm cap}\simeq1.6 \times 10^{-10}M_{\odot}$yr$^{-1}$, and the disk outer radius $R_{\rm d}\simeq 8\times10^{11}$ cm $\simeq 9\times10^5R_{\rm S}$, where $R_{\rm S}=2GM_{\rm co}/c^2$ is the Schwarzschild radius.

For a specific accretor, one can estimate the X-ray luminosity due to accretion. In the next three subsections, we will separately estimate and analyze the three different possible scenarios for the composition of the dark compact companion: 1) a single BH, 2) BNS, and 3) an NS-WD binary.

\subsection{A single BH as the dark compact companion}

For the BH-RG scenario, $R_{\rm d}>R_{\rm ISCO}$, which meets  the requirement for the disk formation, thus an accretion disk would form around the BH. If the accreting gas cools efficiently via radiation, the flow would take a configuration as standard thin disk \citep{shakura73}, and its radiative luminosity is approximated as  $L_{\rm bol}\simeq GM_{\rm co}\dot{M}/R_{\rm co}$, where $R_{\rm co}$ is the compact accretor's radius. Otherwise, inefficient cooling would result in an advection dominated accretion flow with an extremely low radiative luminosity \citep{narayan94,narayan95}. \cite{narayan95} find that an ADAF occurs at low accretion rates, $\dot{M}\leq \alpha^2\dot{M}_{\rm Edd}\sim 0.001-0.1\dot{M}_{\rm Edd}$, where $\alpha \sim 0.03-0.3$ is the viscosity parameter, and $\dot{M}_{\rm Edd}=1.4\times10^{18}\frac{M_{\rm co}}{M_{\odot}}{\rm g}\,{\rm s}^{-1}$ is the Eddington accretion rate. The estimated wind capture rate in V723 Mon is $\dot{M}_{\rm cap}\simeq 2\times 10^{-3}\dot{M}_{\rm Edd}$, suggesting the accretion disk formed around the BH is probably an ADAF.

Note that the accretion rate considered in \cite{narayan94,narayan95} is constant with radius, whereas later theoretical studies and observations suggest otherwise. \cite{blandford99,blandford04} and \cite{begelman12} suggest an inward decrease of the accretion rate in the hot accretion flow, due to the mass loss from the disk in the form of an outflow. This is referred as adiabatic inflow-outflow solution (ADIOS) model. Compared with the constant $\dot{M}(R)$ scenario, the radial density profile for the ADIOS model is flatter. This model is supported by numerous hydrodynamical and magnetohydrodynamical numerical simulations \citep[e.g.,][]{igumenshchev99,igumenshchev00,igumenshchev03,stone99,stone01,mckinney02,mckinney12,narayan12, yuan12a,yuan12b}. In addition, \cite{wang13} and \cite{shi21} find direct observational evidences for the outflow. 

In the presence of outflow, the net mass accretion rate is the difference between inflow and outflow rates. The radial profile of the inflow rate can be described as \citep{yuan12a}
\begin{equation} 
{\dot M_{{\rm{in}}}}\left( R \right) = {\dot M_{{\rm{cap}}}}{\left( {\frac{R}{{{R_{{\rm{out}}}}}}} \right)^s},\, (R>10R_{\rm S})
\end{equation}
where $R_{\rm out}$ represents the outer edge radius of accretion, $s$ is equal to 0.65 and 0.54 for $\alpha$ = 0.001 and 0.01, respectively. If $R<10R_{\rm S}$, the inflow rate keeps constant, i.e., $s=0$, due to the finding by \cite{yuan12a} that the inflow rate is significantly larger than the outflow rate. Adopting an intermediate value of $s=0.4$, we have the accretion rate at $R_{\rm S}$ as
\begin{equation} 
\begin{split}
\dot{M}_{\rm acc}(R_{\rm S})&\simeq {\dot M_{{\rm{cap}}}}{\left( {\frac{10R_{\rm S}}{{{R_{{\rm{d}}}}}}} \right)^{0.4}}\\
&=2\times 10^{-12}M_{\odot}{\rm yr^{-1}}\approx3\times{10}^{-5} \dot M_{\rm Edd}.
\end{split}
\end{equation} 

To get the bolometric luminosity, it is convenient to introduce a radiative efficiency, $\varepsilon$, as in 
\begin{equation}
L_{\rm bol}=\varepsilon \dot{M}_{\rm acc}(R_{\rm S})c^2. 
\end{equation}
Particularly for ADAF, taking into account both the outflow and the direct electron heating by turbulent dissipation, \cite{xie12} systematically studied the radiative efficiency and developed a fitting formula
\begin{equation}
\varepsilon=\varepsilon_0\left[\frac{\dot{M}_{\rm acc}(R_{\rm S})}{0.01\dot{M}_{\rm Edd}}\right]^b,
\end{equation}
where $\varepsilon_0$ and $b$ are the fitting parameters, which both depend on the parameter $\delta$ (ranging from $10^{-3}$ to 0.5) that describes the fraction of turbulent dissipation energy received by the electrons. Note that $L_{\rm bol}$ does not only account for the radiation from the material accreted  at $R_{\rm S}$, but is the total radiation from the whole ADAF. It is worth mentioning that the fitted value of $\varepsilon$ (Eq. 2.10) actually has taken into account the contribution to $L_{\rm bol}$ from larger radii of the disk.

To estimate the bolometric luminosity from the accretor in V723 Mon, we take an intermediate value of $\delta=0.1$, the corresponding $\varepsilon_0$ and $b$ being 0.12 and 0.59 \citep[see][their Tab. 1]{xie12}, respectively, for which Eq. (2.10) gives $\varepsilon= 4\times10^{-3}$. Therefore with Eq. (2.9) we get  $L_{\rm bol}=4\times10^{32}$ erg $\rm s^{-1}$. 

The spectrum from an ADAF usually peaks at frequencies lower than the X-ray band. \cite{niedzwiecki14} calculated ADAF spectra for both stellar-mass ($10M_{\odot}$) and supermassive ($10^{8}M_{\odot}$) black holes, and showed that for the stellar-mass BH, the ratio  $L_{\rm X}$/$L_{\rm bol}$ is around $10^{-2}$ (see their Tab. 1, Fig. 4 and Fig. 9). Adopting this value, one finds that the resulting $L_{\rm X}$ for V723 Mon is around $10^{30}$ erg s$^{-1}$, which is consistent with $L_{\rm X,obs}\approx 8\times 10^{29}$ erg s$^{-1}$. Therefore, the single BH scenario is consistent with the X-ray constraint.

\subsection{Binary neutron stars as the dark companion}\label{sec:redshift}

Next, we consider the slightly less exotic possibility that the dark companion is an inner binary of double NSs. Let $a_{\rm in}$ be the separation of the two NSs. As pointed out by \cite{jaya21}, the inner binary is possibly long-lived and dynamical stable only when $4R_{\odot}<a_{\rm in}<31R_{\odot}$. The two NSs obit each other and accrete the stellar wind material captured within a common cylinder of radius $R_{\rm cap}$. Since $R_{\rm cap}>a_{\rm in}$, the inner binary acts like a single accretor whose mass is equal to the total mass 3$M_{\odot}$. For simplicity, we assume the two neutron stars have the same mass. The total X-ray luminosity is about twice that from a single NS with mass $M_{\rm NS}=1.5M_{\odot}$.

The occurrence of significant accretion requires the inner edge of the accretion disk lie inside the co-rotation radius $R_{\rm cor}$ \citep{pringle72}. The latter is defined as the radius at which a fluid element corotating with the NS experiences a balance between the centrifugal force and the gravity \citep{pringle72}
\begin{equation}
{R_{\rm cor}} = {\left( {\frac{{GM_{\rm NS}}}{{4{\pi ^2}}}} \right)^{{1 \mathord{\left/
 {\vphantom {1 3}} \right.
 \kern-\nulldelimiterspace} 3}}}P_{\rm spin}^{{2 \mathord{\left/
 {\vphantom {2 3}} \right.
 \kern-\nulldelimiterspace} 3}}
\end{equation}
where $P_{\rm spin}$ is the spin period of the neutron star. 

Due to the strong magnetic field of the NS, the inner part of the disk is expected to truncate around the magnetospheric boundary 
\begin{equation}
{R_{\rm m}} = {\left[ {\frac{{R_0^6B_0^2}}{{2{{\dot M}_{\rm in}}(R_{\rm m})\sqrt {2GM_{\rm NS}} }}} \right]^{{2 \mathord{\left/
 {\vphantom {2 7}} \right.
 \kern-\nulldelimiterspace} 7}}},
\end{equation}
at which the ram pressure of the infalling plasma is equal to the magnetic pressure \citep{lamb73,davidson73}, where $R_0$ and $B_0$ are the radius and the surface field of the NS. Because of the low capture rate, here we consider the accretion flow for the NS is similar to that of the BH as discussed in section 2.1. $\dot{M}_{\rm in}(R_{\rm m})$ represents the inflow rate in $R_{\rm m}$, which can be derived from Eq. (2.7).

The fate of accreted material critically depends on the ratio of $R_{\rm m}$ over $R_{\rm cor}$. In the case that $R_{\rm m}>R_{\rm cor}$, which is referred as propeller regime \citep{illarionov75}, the accreted material will be ejected from the system due to the dominance by centrifugal force. This will happen when the NS spins very fast or/and is strongly magnetized. Conversely, if $R_{\rm m}<R_{\rm cor}$, the NS is in the accretion regime. In this case, the matter which has found its way onto field lines just inside $R_{\rm m}$ will flow along the field lines toward the magnetic poles, and is essentially unhindered by the centrifugal force \citep{lamb73}. 

In the propeller regime, most of the infalling matter is propelled away or halted at the boundary $R_{\rm m}$ \citep{campana18}. It is much less efficient in accreting matter than in  the accretion regime. For the latter, as the accreted matter flows along field lines onto the NS, the X-rays would then be emitted thermally from regions near the magnetic poles \citep{pringle72}. 


\begin{figure}
\centering
\includegraphics[width=8.7cm, angle=0]{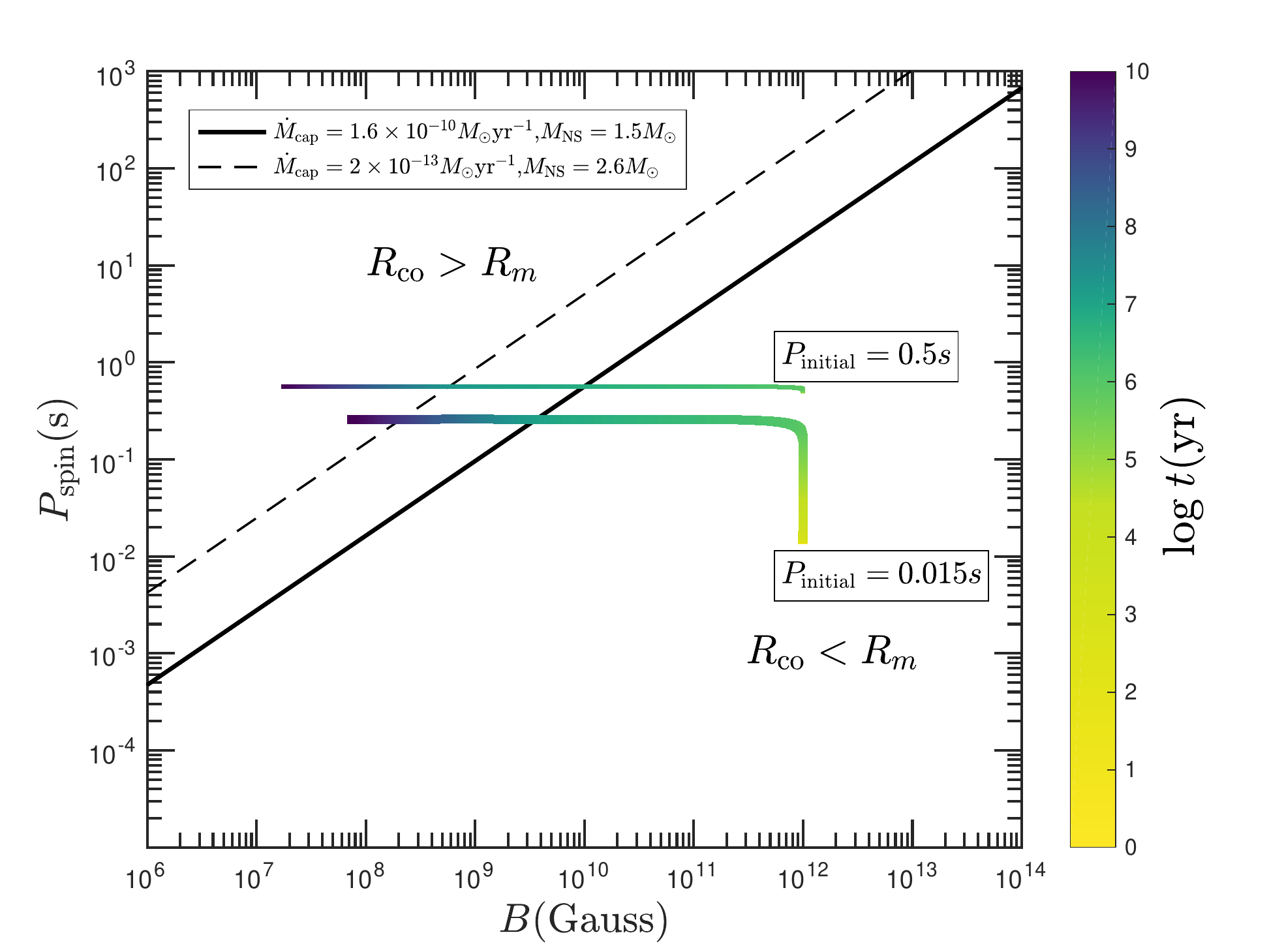}
\caption{The $B-P_{\rm spin}$ parameter space for the accretion regime (top-left) and the propeller regime (bottom-right) of an NS, based on Eqs.(2.11)-(2.12). The two lines are for $R_{\rm cor}=R_{\rm m}$, where the solid line corresponds to $\dot{M}_{\rm cap}=1.6\times10^{-10}M_{\odot}\rm{yr^{-1}}$ and $M_{\rm NS}=1.5M_{\odot}$ for V723 Mon, while the dashed line $\dot{M}_{\rm cap}=2\times10^{-13}M_{\odot}\rm{yr^{-1}}$ and $M_{\rm NS}=2.6M_{\odot}$ for 2MASS J05215658+4359220 (see section 4). The radius of the NS is fixed to be $10^6$ cm. The thick and thin curves with gradient color show the evolution of an NS with initial spin periods of $0.015$ s and $0.5$ s, respectively. The initial magnetic field  is $10^{12}$ G. The two curves are reproduced from \protect\cite{cruces19}, the blue and red solid lines in their Figure 1.}\label{fig:rm_rc}
\end{figure}


Adopting the estimated capture rate, we derive the $B_{\rm 0}-P_{\rm spin}$ parameter space for the propeller vs. accretion regimes in Figure \ref{fig:rm_rc}. The black line in Figure \ref{fig:rm_rc} represents the condition $R_{\rm m}=R_{\rm cor}$. 

Next, we infer the location of the putative NS for V723 Mon in Figure \ref{fig:rm_rc}, i.e., its current $P_{\rm spin}$ and $B_{\rm 0}$. The main-sequence lifetime of the progenitor of the RG V723 Mon is around \citep{kippenhahn90} 
\begin{equation}
{t_{{\rm{MS}}}} \approx {10^{10}}{M_{\rm RG}^{ - 2.5}}{\rm{yr}}.
\end{equation}
This equation roughly holds for main-sequence masses in the range of 0.1-50$M_{\odot}$ \citep{kippenhahn90}. We assume that both the progenitors of the RG and NS were born around the same time. Since the evolutionary time of the NS progenitor is much shorter than that of the RG progenitor, it means that before the initiation of the significant stellar wind accretion from the RG, the neutron star may have enough time ($\sim 10^{10}$ yr) to experience prominent magnetic field decay through ambipolar diffusion \citep{goldreich92} and magnetic dipole spin-down. 

As the ambipolar diffusion is efficient once the NS has cooled down, \cite{cruces19} explored the magnetic ambipolar diffusion in binary systems before the accretion started. They found that a newborn NS with initial field strength $B=10^{12}$ G and initial spin-period $P_{\rm spin}=0.015$ s would evolve to a slower spinning ($\sim0.2$ s) and weakly magnetized ($\sim10^8$ G) NS after $\sim10^9$ yr (see the  thick curve with gradient color in Figure \ref{fig:rm_rc}). The thin curve with gradient color in Figure \ref{fig:rm_rc} shows that a newborn NS with initial $B=10^{12}$ G and slower initial $P_{\rm spin}=0.5$ s would evolve to $10^8$ G and 0.55s within $\sim10^9$ yr, respectively. These two curves are reproduced directly from \cite{cruces19}. Note that these initial $B$ and $P_{\rm spin}$ are consistent with the typical values of the inferred newborn NS, i.e., $10^{13}$ G and 100 ms \citep{faucher06,gullon14,ho20}. Therefore, it is very probably that the NS has possessed a weak magnetic field and rotated slowly before the accretion begins, which would put the current NS in the accretion regime.

At low accretion rate $\approx \dot{M}_{\rm cap}\sim10^{-10}M_{\odot}$yr$^{-1}$, the accretion disk of the NS is likely advection dominated. However, the hard surface of the NS would reradiate all the advected energy \citep{narayan94,mahadevan97}, thus the resulting X-ray luminosity of the NS can be roughly estimated as
\begin{equation}
\begin{split}
{L_{\rm X}} &\simeq \frac{{G{M_{\rm NS}}{{\dot M}_{\rm cap}}}}{{{R_{\rm NS}}}} = 2 \times {10^{36}}{\rm{erg}}\,{{\rm{s}}^{-1}}\left(\frac{{{M_{\rm NS}}}}{{1.5{M_ \odot }}}\right)\\
 &\times \left(\frac{{{{\dot M}_{\rm cap}}}}{{1.6 \times {{10}^{ - 10}}{M_ \odot }{\rm{y}}{{\rm{r}}^{ - 1}}}}\right)\left(\frac{{{R_{\rm NS}}}}{{{{10}^{6}}{\rm{cm}}}}\right)^{ - 1}
\end{split}
\end{equation}
which exceeds $L_{\rm X, obs}$ by 6 orders of magnitude. Thus, this scenario is unlikely.

On the other hand, one may have reasons, though not strong, to still consider the possibility that the NS is in the propeller regime. For instance, the putative triple system in V723 Mon might have formed dynamically in dense stellar environments in which the NS was captured, so the NS is too young to undergo significant field decay. Besides, the decay of magnetic field may not have resulted from ambipolar diffusion, but through ohmic dissipation or through screening by the accreted matter \citep{cruces19,igoshev21}, so that the field strength may not be weak enough. However, theoretical calculation \citep{stella86} and magnetohydrodynamic simulations \citep[e.g.,][]{romanova04,ustyugova06} suggest that even the accretion is inhibited in the equatorial plane of the NS, a small fraction ($\sim0.01$) of matter near $R_{\rm m}$ at high latitudes may penetrate the magnetosphere and reach the surface of the NS. Some observations suggest that the transition from accretion regime into propeller regime, e.g., 4U 0115+63 and V0332+53 in high-mass X-ray binaries \citep{stella86,tsygankov16}, SAX J1808.4--3658 \citep{campana08} and XTE J1701--462 \citep{fridriksson10} in low-mass X-ray binaries, causes a drop in $L_{\rm X}$ by a factor $\sim1000$. Thus considering these, even if the NS may stay in propeller regime, the resulting $L_{\rm X}$ could be reduced by a factor of $10^3$, i.e., down to $10^{33}$ erg s$^{-1}$, which is still much larger than $L_{\rm X, obs}$.

\subsection{NS-WD binary as the dark companion}\label{sec:mw-archival}

In the third scenario that the dark companion is an inner binary of an NS and a WD, due to the presence of an NS, the inferred X-ray luminosity from the NS-WD binary is at least about half of that in the BNS scenario. Therefore, similar to the BNS scenario (section 2.2), this scenario is also unlikely.



\section{Emission from the captured stellar wind}\label{sec:vhe_prospects}

As the compact star moves supersonically through the wind from the RG, a bow shock may form and produce synchrotron radiation detectable at radio and infrared frequencies \citep{wang14,ginsburg16}. \cite{hunt71} has shown the presence of bow shocks when a gravitating point source is moving supersonically through an adiabatic gas (with a specific heat ratio 5/3) for small Mach numbers. In high-resolution 3D simulations, \cite{huarte13} revealed the presence of bow shocks around accretion disks of a stellar mass secondary orbiting a wind-producing asymptotic giant branch (AGB) primary. For the V723 Mon system, the wind velocity relative to the compact object is around 120 km/s, which is larger than the local sound speed ($\sim 20$ km/s). Therefore, a bow shock probably forms. In this section, we adopt the method of \cite{wang14} to estimate the synchrotron emission from the accelerated electrons in the bow shock. 

Assuming energy equipartition, the post-shock magnetic field can be estimated as \citep{chevalier98,sari98}
\begin{equation}
B=\sqrt{4\pi\epsilon_{\rm B}\rho_{\rm w}v_{\rm rel}^2}
\end{equation}
where $\epsilon_{\rm B}$ is the density ratio of the magnetic energy to the internal energy of the post shock fluid.

The total emitted power per unit frequency from a single electron with Lorentz factor $\gamma$ is given by \citep{rybicki79},
\begin{equation}
P\left( \nu  \right) = \sqrt 3 \frac{{{e^3}B}}{{{m_{\rm e}}{c^2}}}F\left( {\frac{\nu }{{{\nu _{\rm c}}}}} \right)
\end{equation}
where $F\left(x\right) = x\int_x^\infty  {{K_{{5 \mathord{\left/
 {\vphantom {5 3}} \right.
 \kern-\nulldelimiterspace} 3}}}\left( \xi  \right)} {\rm{d}}\xi$, $K_{5/3}$ is the modified Bessel function of the second kind of 5/3 order, $c$ is the speed of light and $m_{\rm e}$ is the electron mass, $e$ is the electron charge, and ${\nu _{\rm c}} = \frac{{3{\gamma ^2}eB}}{{4\pi {m_{\rm e}}c}}$ is the critical frequency. The total synchrotron emission power from a single electron, which has an isotropic distribution of velocities, is given by \citep{rybicki79}
\begin{equation}
P_{\rm syn} = \frac{4}{9} r_{\rm e}^2c{\beta ^2}{\gamma ^2}{B^2}
\end{equation}
where $r_{\rm e}=e^2/m_{\rm e}c^2$ is the classical radius of an electron.

According to numerical models of shock acceleration \citep[e.g.,][]{park15,crumley19}, the post-shock electrons mainly reside in a thermal population and carry a dominant fraction of the total post-shock energy over that of the non-thermal electrons, which has been applied to explain the steep optically-thin spectral of AT2018cow and AT2020xnd \citep{ho21}. \cite{margalit21} establish a model of `thermal + non-thermal' synchrotron radiation from sub-relativistic shocks and derive properties of the resulting spectrum and light curves. The prominent properties of the thermal electron synchrotron emission is a steep optically-thin spectral index and a largely optically thick spectrum, where $F_{\rm \nu}\propto\nu^2$ \citep{margalit21}. Motivated by those works, we assume that a small fraction of the electrons, $\epsilon_{\rm n}$, are non-thermal, and the others are thermal and occupy a Maxwell-J$\rm\ddot{u}$ttner distribution. Thus the number density of the non-thermal electrons is $n_{\rm e}\epsilon_n$, where $n_{\rm e}=\rho_{\rm s}/m_{\rm p}$, $m_{\rm p}$ is the mass of proton and $\rho_{\rm s}=4\rho_{\rm w}$ is the density of the shocked gas resulting from a constant shock compression ratio of 4.

Assuming that the non-thermal electrons have a broken power-law distribution due to synchrotron cooling, the number density of particles with Lorentz factor between $\gamma$ and $\gamma+d\gamma$ is \citep{wang14}
\begin{equation}
N\left( \gamma  \right)d\gamma = \begin{array}{*{20}{c}}
{{N_0}{\gamma ^{ - p}}{{\left( {1 + \frac{\gamma }{{{\gamma _{\rm b}}}}} \right)}^{ - 1}}}d\gamma,&{ {{\gamma _{\min }} \le \gamma  \le {\gamma _{\max }}}}
\end{array}
\end{equation}
where $N_0$ is the normalization factor, whose value is determined by 
\begin{equation}
n_{\rm e}\epsilon_{\rm n} = \int_{{\gamma _{\min }}}^{{\gamma _{\max }}} {{{N_0}{\gamma ^{ - p}}{{\left( {1 + \frac{\gamma }{{{\gamma _{\rm b}}}}} \right)}^{ - 1}}}} {\rm{d}}\gamma,
\end{equation}
$p$ is the power law index, and $\gamma_{\rm b}$ is the break (due to cooling) Lorentz factor. With an assumption that a fraction, $\epsilon_{\rm e}$, of the shock kinetic energy goes into accelerated non-thermal electrons, the minimum Lorentz is given by 
\begin{equation}
\gamma_{\rm min}-1\approx\epsilon_{\rm e}\frac{m_{\rm p}}{m_{\rm e}\epsilon_{\rm n}}\frac{v_{\rm rel}^2}{c^2}
\end{equation}
which is in accordance with \cite{ho19} except for the factor $\epsilon_{\rm n}$ in the denominator for the reason that only non-thermal electrons occupy the power-law distribution. The maximum Lorentz factor $\gamma_{\rm max}$ is set by the constraint that electron acceleration timescale does not exceed  the cooling timescale or the dynamical timescale \citep{wang14}.  

The synchrotron elections cool down on a time scale of \citep{wang14}
\begin{equation}
\begin{split}
{t_{\rm cool}} &= {{\gamma {m_{\rm e}}{c^2}} \mathord{\left/
 {\vphantom {{\gamma {m_{\rm e}}{c^2}} {{P_{\rm syn}}}}} \right.
 \kern-\nulldelimiterspace} {{P_{\rm syn}}}}\\
 &= 25{\left( {\frac{\gamma }{{10^4}}} \right)^{ - 1}}{\left( {\frac{B}{{0.01\;{\rm{G}}}}} \right)^{ - 2}}{\rm{yr}}.
\end{split}
\end{equation}
The dynamical time scale derives from the fact that most of the synchrotron emission is generated around the head of the Mach core. So it is given by \citep{wang14}
\begin{equation}
\begin{split}
{t_{\rm dyn}} &= {{{R_{\rm cap}}} \mathord{\left/
 {\vphantom {{{R_{\rm cap}}} {{v_{\rm rel}}}}} \right.
 \kern-\nulldelimiterspace} {{v_{\rm rel}}}}\\
 &= 1.3\times{10^{-2}}\left( {\frac{{{R_{\rm cap}}}}{{5 \times {{10}^{12}}\;{\rm{cm}}}}} \right){\left( {\frac{{{v_{\rm rel}}}}{{1.2 \times {{10}^7}{{{\rm{cm}}} \mathord{\left/
 {\vphantom {{{\rm{cm}}} {\rm{s}}}} \right.
 \kern-\nulldelimiterspace} {\rm{s}}}}}} \right)^{ - 1}}{\rm{yr}}.
\end{split}
\end{equation}
As $t_{\rm dyn}$ is smaller than $t_{\rm cool}$, $\gamma_{\rm max}$ should be determined by equating the electron acceleration time scale ${t_{\rm acc}} = {{{\xi _{\rm acc}}\gamma {m_{\rm e}}{c^3}} \mathord{\left/
 {\vphantom {{{\xi _{\rm acc}}\gamma {m_{\rm e}}{c^3}} {eBv_{\rm rel}^2}}} \right.
 \kern-\nulldelimiterspace} {eBv_{\rm rel}^2}}$ \citep{blandford87}, where $\xi_{\rm acc}$ is a dimensionless constant of order unity, with the dynamical timescale $t_{\rm dyn}$, as
\begin{equation}
\begin{split}
{\gamma _{\max }} &= \frac{{eB{v_{\rm rel}}{R_{\rm acc}}}}{{{\xi _{\rm acc}}{m_{\rm e}}{c^3}}}\\
&=1.2\times 10^4 (\frac{B}{0.01G})(\frac{v_{\rm rel}}{1.2\times 10^7 {\rm cm/s}})(\frac{R_{\rm acc}}{5\times 10^{12} {\rm cm}})
\end{split}
\end{equation}
The cooling break Lorentz factor is given by equating $t_{\rm cool}$ with $t_{\rm dyn}$, so
\begin{equation}
{\gamma _{\rm b}} = 1.6 \times {10^7}{\left( {\frac{{{v_{\rm rel}}}}{{1.2 \times {{10}^7}{{{\rm{cm}}} \mathord{\left/
 {\vphantom {{{\rm{cm}}} {\rm{s}}}} \right.
 \kern-\nulldelimiterspace} {\rm{s}}}}}} \right)^3}{\left( {\frac{B}{{0.01\;{\rm{G}}}}} \right)^{ - 2}}{\left( {\frac{M}{{3{M_ \odot }}}} \right)^{ - 1}}.
\end{equation}
Only the electrons with Lorentz factor larger than $\gamma_{\rm b}$ could cool and stop emitting synchrotron radiation within $t_{\rm dyn}$. As $\gamma_{\rm b} \gg \gamma_{\rm max}$, the synchrotron cooling is inefficient. 

\begin{figure}
\begin{center}
\includegraphics[width=8.7cm, angle=0]{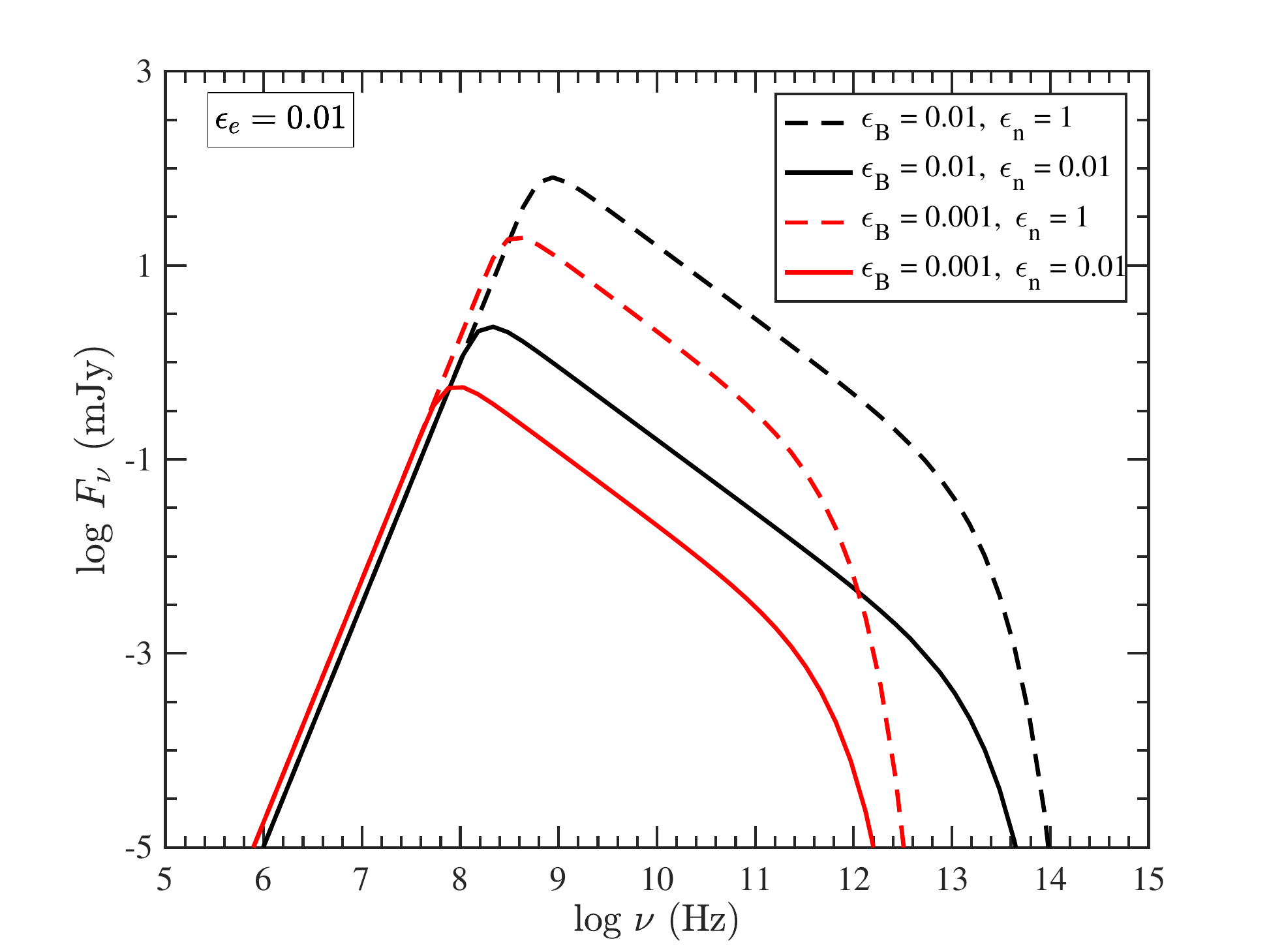}

\caption{The non-thermal electrons producing synchrotron flux observed on the earth. The slope of the electron distribution is set to 2.5 (see Eq. 3.4). The distance between the source and the earth is 460 pc. The size of the source is approximated as $R_{\rm acc}$. The values of $\epsilon_{\rm n}$ are set to 0.01 (solid) and 1 (dashed). Compared with $\epsilon_{\rm n}=1$, $\epsilon_{\rm n}=0.01$ means that the fraction of non-thermal electrons is reduced by a factor of 100, resulting in a smaller flux by around 2 orders of magnitude. The red and black lines are shown for $\epsilon_{\rm B}$ = 0.001 and 0.01, respectively.}\label{fig:F_nu_p2_5_etaB}
\end{center}
\end{figure}

The emissivity and self-absorption coefficients for synchrotron radiation are given as \citep{rybicki79}
\begin{equation}
\begin{split}
{j_{\rm \nu} } &= \int_{{\gamma _{\rm min }}}^{{\gamma _{\rm max }}} {P\left( \nu  \right)N\left( \gamma  \right)} {\rm{d}}\gamma \\
 &= \frac{{\sqrt 3 {e^3}B}}{{{m_{\rm e}}{c^2}}}\int_{{\gamma _{\rm min }}}^{{\gamma _{\rm max }}} {F\left( {\frac{\nu }{{{\nu _{\rm c}}}}} \right)N\left( \gamma  \right)} {\rm{d}}\gamma 
\end{split}
\end{equation}
and 
\begin{equation}
\begin{split}
{\alpha _\nu } &= \int_{{\gamma _{\rm min }}}^{{\gamma _{\rm max }}} {P\left( \nu  \right){\gamma ^2}\frac{{\rm{d}}}{{{\rm{d}}\gamma }}\left( {\frac{{N\left( \gamma  \right)}}{{{\gamma ^2}}}} \right)} {\rm{d}}\gamma \\
 &= \frac{{\sqrt 3 {e^3}B}}{{{m_{\rm e}}{c^2}}}\int_{{\gamma _{\rm min }}}^{{\gamma _{\rm max }}} {F\left( {\frac{\nu }{{{\nu _{\rm c}}}}} \right){\gamma ^2}\frac{{\rm{d}}}{{{\rm{d}}\gamma }}\left( {\frac{{N\left( \gamma  \right)}}{{{\gamma ^2}}}} \right)} {\rm{d}}\gamma, 
\end{split}
\end{equation}
respectively. The observed flux density is given by \citep{rybicki79}
\begin{equation}
{F_\nu } = \frac{{\pi R_{\rm acc}^2}}{{{D^2}}}\frac{{{j_\nu }}}{{{\alpha _\nu }}}\left[{1 - \exp \left( { - {\tau _\nu }} \right)} \right]
\end{equation}
where the optical depth is estimated as $\tau_{\rm \nu}\approx R_{\rm cap}\alpha_{\nu}$, and $D$ is the distance of the source.

The synchrotron spectra calculated as above are shown in Figure \ref{fig:F_nu_p2_5_etaB}. We use a standard value $p=2.5$ \citep{sari96,sari98}. We set  the fraction $\epsilon_{\rm n}$ to 0.01 (solid) and 1 (dashed), respectively, for comparison. The red and black lines are shown for $\epsilon_{\rm B}$ = 0.001 and 0.01, respectively. According to Eq. (3.6), as the shock velocity is around $10^{-3}$c, $\gamma_{\rm min}$ is about 1 and is insensitive to $\epsilon_{\rm e}$. Thus we set $\epsilon_{\rm e}=0.01$.   

As shown in Figure \ref{fig:F_nu_p2_5_etaB}, there is a low-frequency break around $10^8$ ($10^9$) Hz for $\epsilon_{\rm n}=$ 0.01 (1), where $\tau_{\rm \nu}=1$, below which $F_{\nu} \propto \nu^{5/2}$. The high-frequency break is determined by $\gamma_{\rm max}$, and $B$ which is dependent on $\epsilon_{\rm B}$. The region between the two brakes corresponds to the slow-cooling, optical-thin regime, where $F_{\rm \nu}\propto \nu^{-(p-1)/2}$. 

The synchrotron emissions peak at a frequency $\sim$ 0.1--1 GHz, with a flux density $\sim$ 1--10 mJy. It is observable for the Very Large Array Sky Survey \citep[VLASS,][]{lacy20} whose detection limit is $\sim$ 70 $\mu$Jy for 1$\sigma$ detection at 2-4 GHz. The Five-hundred-metre Aperture Spherical Telescope \citep[FAST,][]{li18}, operating at frequencies from 70 to 3000 MHz and having a minimum detectable flux density down to 0.6$\mu$Jy for pulsars \citep{smit09}, is also able to detect this emission.  


It is possible to distinguish the bow shock emission from the radio radiation of the ADAF disk because of the difference in their luminosities. \cite{mahadevan97} presented several scaling laws for ADAFs, by considering cooling of electrons through synchrotron, bremsstrahlung, and Compton processes. We estimate the possible synchrotron radio emission from the ADAF in V723 Mon via the scaling law given by \cite{mahadevan97} as
\begin{equation}
\begin{split}
&{L_\nu } = 5\times10^{15}{\left( {\frac{\alpha }{{0.1}}} \right)^{{{ - 4} \mathord{\left/
 {\vphantom {{ - 4} 5}} \right.
 \kern-\nulldelimiterspace} 5}}}{\left( {\frac{\nu }{{{{10}^9}{\rm{Hz}}}}} \right)^{{2 \mathord{\left/
 {\vphantom {2 5}} \right.
 \kern-\nulldelimiterspace} 5}}} {\left( {\frac{{{T_{\rm e}}}}{{2 \times {{10}^9}{\rm{K}}}}} \right)^{{{21} \mathord{\left/
 {\vphantom {{21} 5}} \right.
 \kern-\nulldelimiterspace} 5}}}\\
&\times{\left( {\frac{{{M_{\rm BH}}}}{{3{M_ \odot }}}} \right)^{{6 \mathord{\left/
 {\vphantom {6 5}} \right.
 \kern-\nulldelimiterspace} 5}}}{\left( {\frac{{{{\dot M}_{\rm cap}}}}{{1.6 \times {{10}^{ - 10}}{M_ \odot }{\rm{y}}{{\rm{r}}^{ - 1}}}}} \right)^{{4 \mathord{\left/
 {\vphantom {4 5}} \right.
 \kern-\nulldelimiterspace} 5}}}{\rm{erg}}\,{{\rm{s}}^{ - 1}}{\rm{H}}{{\rm{z}}^{ - 1}}
\end{split}
\end{equation}
where $T_{\rm e}$ is the equilibrium temperature of the electrons. Note that this equation does not take the outflow into account. Thus the flux density $F_{\rm \nu,1GHz}=L_{\rm \nu}/4\pi D^2\simeq 20$ $\mu$Jy, which is smaller than that of the bow shock by $\sim 1$ order of magnitude (see Figure. \ref{fig:F_nu_p2_5_etaB}).

We attempt to search possible candidate detection of V723 Mon in radio source catalogs provided by High Energy Astrophysics Science Archival Research Center (HEASARC). We find a closest radio source PMN J0628-0536 with J2000 coordinates ($\alpha$, $\delta$)=(06 28 32.10, -05 36 30.0), at an 8.385$'$ offset towards V723 Mon, found in Parkes-MIT-NRAO (PMN) Surveys \citep{griffith95}, with a flux density of 106$\pm 12$ mJy at 4850 MHz. It is much brighter than the bow shock synchrotron emission ($\sim$1 mJy at 4850 MHz, for $\epsilon_{\rm B}$ = 0.01) and even so than the synchrotron cooling emission ($3.8\times10^{-2}$ mJy at 4850 MHz) of the ADAF disk. Therefore, the relatively large offset and the discrepancy between the observed and model-predicted flux densities disfavor the hypothesis that PWN J0628-0536 is physically related to V723 Mon.

\section{Summary and Discussion}\label{sec:discussion}

Identifying the nature of the non-interacting dark compact objects in binaries or triples around luminous companions is important  to constrain the pathways of forming them, and characterizing NS and BH is crucial to the understanding of deaths of massive stars and core-collapse supernovae \citep{thompson19,jaya21}. V723 Mon is reported to have a dark companion with a mass of $\sim3M_{\odot}$ \citep{jaya21,masuda21}. So far, only one non-interacting mass-gap BH candidate has been found \citep{thompson19}. 

In this paper, with the aim to reveal the nature of the dark companion, we consider the mass loss from the RG as the origin of the wind-fed accretion onto the dark companion and compare the estimated X-ray emission with the observation. We analyze three different scenarios, i.e., a single BH, binary neutron stars and binary of a WD and an NS as the dark companion. Besides, we estimate  the synchrotron radio emission from the bow shock as the signature of the wind-fed accretion.

We find that the two scenarios involving an inner binary as the dark companion are unlikely, due to their overproduction of the X-ray luminosity. While for the single BH scenario, as the accretion disk is an ADAF due to the low accretion rate and most of the energy disappear in the horizon of the BH, the corresponding luminosity is relatively low, $L_{\rm X}\approx 10^{29-30}$ erg s$^{-1}$, which is compatible with $L_{\rm X,obs}\sim 7\times10^{29}$ erg $\rm s^{-1}$.

We show that the synchrotron emissions from the bow shock may peak at the frequency $\sim$ 0.1--1 GHz, with a flux density around 1 mJy, which is detectable for VLASS and FAST. This signal would be periodic for the orbiting of the BH around the RG, providing an evidence for the wind-fed accretion and a method to find the non-interacting dark compact object. 

Here, we also apply our approach to the non-interacting binary system consisted of a rapidly rotating giant star 2MASS J05215658+4359220 with mass of $3.2_{-1.0}^{+1.0}$ $M_{\odot}$ and a massive unseen companion of $3.3_{-0.7}^{+2.8}$ $M_{\odot}$ \citep{thompson19}. Non-detection by Swift X-ray Telescope and Ultraviolet/Optical Telescope provides an upper limit of $L_{\rm X}\leq10^{31}$ erg s$^{-1}$. \cite{thompson19} excluded the possibility that the unseen companion is a massive NS. This conclusion is in accordance with our quantitative analysis. Specifically, if it was an NS with mass of $2.6M_{\odot}$ (i.e., the lower limit of the massive unseen object), for the mass capture rate $\sim 2 \times 10^{-13} M_{\odot} \rm{yr^{-1}}$ \citep{thompson19}, the NS needs to spend about $10^{7.5}$ years (estimated from the intersection point of the dashed line and the thin curve with gradient color shown in Figure \ref{fig:rm_rc}) to have a weak magnetic field ($\sim10^8$ G) and rotate slowly. This timescale is shorter than the main-sequence lifetime ($\approx 5\times10^8$ yr) of the giant star according to Eq. (2.13). Thus the NS may already be in the accretion state, such that the resulting $L_{\rm X}\simeq GM\dot{M}_{\rm cap}/R_{\rm NS}\sim10^{33}$ erg s$^{-1}$ would exceed the above upper limit by 2 orders of magnitude. 


\section*{Acknowledgements}

Y.L. and R.-F.S. are supported by the National Natural Science Foundation of China (12073091), China Manned Spaced Project (CMS-CSST-2021-B11) and Guangdong Basic and Applied Basic Research Foundation (2019A1515011119). E.L.Q. is supported by the National Natural Science Foundation of China (Grants 12173048) and NAOC Nebula Talents Program.

\section*{Data availability}

The data underlying this article will be shared on reasonable request to the corresponding author.


\bsp	
\label{lastpage}
\end{document}